\documentclass[12pt]{article}
\usepackage{a4}
\usepackage{graphics}

\makeatletter

\newcommand{\LyX}{L\kern-.1667em\lower.25em\hbox{Y}\kern-.125emX\@}

\makeatother

\begin{document}

\title{Mean field and Monte Carlo studies of the magnetization-reversal transition
in the Ising model }

\maketitle
{\par\centering Arkajyoti Misra\footnote{
email : arko@cmp.saha.ernet.in
} and Bikas K Chakrabarti\footnote{
email : bikas@cmp.saha.ernet.in
}\par}

{\par\centering Saha Institute of Nuclear Physics, 1/AF Bidhannagar, Calcutta
700 064, India.\par}

\begin{abstract}
Detailed\( \:  \) mean field and Monte Carlo studies of the dynamic magnetization-reversal
transition in the Ising model in its ordered phase under a competing external
magnetic field of finite duration have been presented here. Approximate analytical
treatment of the mean field equations of motion shows the existence of diverging
length and time scales across this dynamic transition phase boundary. These
are also supported by numerical solutions of the complete mean field equations
of motion and the Monte Carlo study of the system evolving under Glauber dynamics
in both two and three dimensions. Classical nucleation theory predicts different
mechanisms of domain growth in two regimes marked by the strength of the external
field, and the nature of the Monte Carlo phase boundary can be comprehended
satisfactorily using the theory. The order of the transition changes from a
continuous to a discontinuous one as one crosses over from coalescence regime
(stronger field) to nucleation regime (weaker field). Finite size scaling theory
can be applied in the coalescence regime, where the best fit estimates of the
critical exponents are obtained for two and three dimensions.\newpage

\end{abstract}

\section{Introduction}

The study of the response of pure Ising systems under the action of a time-dependent
external magnetic field has been of recent interest in statistical physics \cite{ca99}\cite{srn98.a}\cite{mc98}.
A whole class of dynamic phase transitions emerged from the study of such driven
spin systems under different time dependences of the driving field. A mean field
study was initially proposed by Tome and Oliveira \cite{to90} where the time
dependence of the external perturbation was periodic. Subsequently, through
extensive Monte Carlo studies, the existence of a dynamic phase transition under
periodic magnetic field was established and properly characterized \cite{ac95}\cite{srn98.b}\cite{srn99}.
Later, efforts were made to investigate the response of such systems under magnetic
fields which are of the form of a `pulse' or in other words applied for a finite
duration of time. All the studies with pulsed fields were made on a system below
its static critical temperature \( T_{c}^{0} \), where the equilibrium state
has got a prevalent order along a particular direction. The pulse is called
`positive' when it is applied along the direction of the prevalent order and
`negative' when applied in opposition. The results for the positive pulse case
was analyzed by extending appropriately the finite size scaling technique to
this finite time window case, and it did not involve any new phase transition
or introduced any new thermodynamic scale \cite{abc97}. However a negative
field competes with the existing order and depending on the strength \( h_{p} \)
and duration \( \Delta t \) of the pulse, the system may show a transition
from one ordered state with equilibrium magnetization \( +m_{0} \) (say) to
the other equivalent ordered state with equilibrium magnetization \( -m_{0} \)
\cite{mc97}. This transition is called here the ``magnetization-reversal''
transition. It may be noted that a magnetization-reversal phenomenon trivially
occurs in the limit \( \Delta t\rightarrow \infty  \) for any non vanishing
value of \( h_{p} \) at any \( T<T_{c}^{0} \). However, this is a limiting
case of the transition, which is studied here only for finite \( \Delta t \).
In our studies the magnetization-reversal need not occur during the presence
of the external field. In fact, it will be shown later that the closer one approaches
to the threshold value \( h_{p}^{c} \) of the pulse strength longer is the
time taken by the system, after the field is withdrawn, to relax to the final
ordered state. We report here in details the various results obtained for this
dynamic magnetization-reversal transition in pure Ising model in two and three
dimensions.

The model we studied here is the Ising model with nearest neighbour interaction
under a time dependent external magnetic field, described by the Hamiltonian

\begin{equation}
\label{hamil}
H=-\frac{1}{2}\sum _{\left[ ij\right] }J_{ij}S_{i}S_{j}-\sum _{i}h_{i}(t)S_{i},
\end{equation}
 where \( J_{ij} \) is the cooperative interaction between the spins at sites
\( i \) and \( j \) respectively and each nearest-neighbour pair denoted by
\( \left[ \ldots \right]  \) is counted twice in the summation. We consider
the system at temperatures only below its static critical temperature (\( T<T_{c}^{0} \)).
The external field is applied after the system is brought to equilibrium characterized
by an equilibrium magnetization \( m_{0}(T) \). The field is uniform in space
(\( h_{i}(t)=h(t)\textrm{ for all }i \)) and its time dependence is given by

\begin{equation}
\label{pulse}
\begin{array}{ccccl}
h(t) & = & -h_{p} & , & \textrm{for }t_{0}\leq t\leq t_{0}+\Delta t\\
 & = & 0 & , & \textrm{otherwise}.
\end{array}
\end{equation}
Typical responses of the time dependent magnetization \( m(t) \) under different
\( h(t) \) are shown in figure 1. As mentioned before, for appropriate combination
of \( h_{p} \) and \( \Delta t \), magnetization-reversal transition occurs
when the system makes a transition from one ordered state to another. This transition
can be observed at any dimension \( d \) greater than unity for systems with
short range interactions. This is because one has to work at temperatures \( T<T_{c}^{0} \)
where, in absence of a symmetry breaking field, the free energy landscape has
got two equivalent minima at magnetizations \( m=\pm m_{0} \). A phase boundary
in the \( h_{p}-\Delta t \) plane gives the minimal combination of the two
parameters at a particular temperature \( T \)(\( <T_{c}^{0} \)) required
to bring about the transition.

A full numerical solution as well as an analytical treatment in the linear limit
of the dynamic mean field equation of motion shows the existence of length and
time scale divergences at the transition phase boundary \cite{smc99}. The divergence
of length and time scales is also observed in Monte Carlo (MC) simulation study
of Ising model with nearest neighbour interaction evolving under a negative
pulse through single spin flip Glauber dynamics \cite{mc98}. The phase diagram
for the transition was obtained for both MF and MC studies. While the phase
boundaries for the two cases are qualitatively of similar nature, there exists
a major difference which can be accounted for by considering the presence of
fluctuations in the simulations. In the MC study, there exists two distinct
time scales in the problem : (i) the nucleation time \( \tau _{N} \) is the
time taken by the system to leave the metastable state under the influence of
the external magnetic field and (ii) the relaxation time \( \tau _{R} \) is
the time taken by the system to reach the final equilibrium state after the
external field is withdrawn. While \( \tau _{N} \) is controlled by the strength
\( h_{p} \) of the external pulse and is bounded by its duration \( \Delta t \)
which is finite, \( \tau _{R} \) is the time scale that diverges at the magnetization-reversal
phase boundary. According to the classical nucleation theory (CNT) \cite{gd83},
there can be two distinct mechanisms for the growth of domains or droplets depending
on the strength of the external field. Under the influence of weaker external
magnetic fields, only a single droplet grows to span the entire system and this
is called the single-droplet (SD) or the nucleation regime. On the other hand,
under stronger magnetic fields, many small droplets can grow simultaneously
and eventually coalesce to form a system spanning droplet. This is called the
multi-droplet (MD) or the coalescence regime. The crossover from SD to MD regime
takes place at the dynamic spinodal field or \( h_{DSP}(L,T) \) which is a
function of system size \( L \) and temperature \( T \). The nucleation time
\( \tau _{N} \) changes abruptly as one crosses over from SD to MD regime even
along the same phase boundary. The nature of the transition too changes from
a continuous one in the MD regime to a discontinuous nature in the SD regime.
All our simulation observations for the dynamic phase boundary compare well
with those suggested by the CNT. The investigations about the relaxation time
\( \tau _{R} \) and the correlation length \( \xi  \) are also discussed here.
The application of scaling theory in the MD regime gives the estimates of the
critical exponents for this dynamic transition. The organization of the paper
in as follows : We discuss the MF results in the next section and the MC results
for square and simple cubic lattices in section 3. A brief summary and concluding
remarks are given in section 4.

\section{Mean field study}

The master equation for a system of \( N \) Ising spins in contact with a heat
bath evolving under Glauber single spin flip dynamics can be written as \cite{sk68}

\begin{eqnarray}
\frac{d}{dt}P\left( S_{1},\cdots ,S_{N};\, t\right)  & = & -\sum _{j}W_{j}(S_{j})P\left( S_{1},\ldots ,S_{N};\, t\right) \nonumber \\
 & + & \sum _{j}W_{j}(-S_{j})P\left( S_{1},\ldots ,-S_{j},\ldots ,S_{N};\, t\right) ,\label{master} 
\end{eqnarray}
 where \( P\left( S_{1},\cdots ,S_{N};\, t\right)  \) is the probability to
find the spins in the configuration \( \left( S_{1},\cdots ,S_{N}\right)  \)
at time \( t \) and \( W_{j}(S_{j}) \) is the probability of flipping of the
\( j \)th spin. Satisfying the condition of detailed balance one can write
the transition probability as

\begin{equation}
\label{tran-prob}
W_{j}(S_{j})=\frac{1}{2\lambda }\left[ 1-S_{j}\tanh \left( \frac{\sum _{i}J_{ij}S_{i}(t)+h_{j}}{T}\right) \right] ,
\end{equation}
 where \( \lambda  \) is a temperature dependent constant. Defining the spin
expectation value as

\begin{equation}
\label{m=<s>}
m_{i}=\left\langle S_{i}\right\rangle =\sum _{\{S\}}S_{i}P\left( S_{1},\ldots ,S_{N};\, t\right) ,
\end{equation}
 where the summation is carried over all possible spin configurations, one can
write 

\begin{equation}
\label{d<s>/dt}
\lambda \frac{dm_{i}}{dt}=-m_{i}+\left\langle \tanh \left( \frac{\sum _{j}J_{ij}S_{j}+h_{i}}{T}\right) \right\rangle .
\end{equation}
 Under the mean field approximation (\ref{d<s>/dt}) can be written after a
Fourier transform as

\begin{equation}
\label{dmq/dt}
\lambda \frac{dm_{q}(t)}{dt}=-m_{q}(t)+\tanh \left( \frac{J(q)m_{q}(t)+h_{q}(t)}{T}\right) ,
\end{equation}
 where \( J(q) \) is the Fourier transform of \( J_{ij} \). Equation (\ref{dmq/dt})
is not analytically tractable and one can only look for solutions in the small
\( m_{q} \) limit where terms linear in \( m_{q} \) are dominant. The linearized
equation of motion, therefore, can be written as

\begin{equation}
\label{lin dmq/dt}
\frac{dm_{q}(t)}{dt}=\lambda ^{-1}\left[ \left( K(q)-1\right) m_{q}(t)+\frac{h_{q}(t)}{T}\right] ,
\end{equation}
 where \( K(q)=J(q)/T \). When we are concerned only with the homogeneous magnetization,
we consider the \( q=0 \) mode of the equation and writing \( m_{q=0}=m \)
and \( h_{q=0}=h \), we get

\begin{equation}
\label{lin dm/dt}
\frac{dm}{dt}=\lambda ^{-1}\left[ \left( K(0)-1\right) m(t)+\frac{h(t)}{T}\right] .
\end{equation}
 In the mean field approximation \( K(0)=T_{c}^{MF}/T \) with \( T_{c}^{MF}=J(0) \)
and for small \( q \), \( K(q)\simeq K(0)\left( 1-q^{2}\right)  \). Differentiating
(\ref{dmq/dt}) with respect to the external field, we get the rate equation
for the dynamic susceptibility \( \chi _{q}(t) \) as

\begin{equation}
\label{dxq/dt}
\lambda \frac{d\chi _{q}(t)}{dt}=-\chi _{q}(t)+\left( \frac{J(q)\chi _{q}(t)+1}{T}\right) \textrm{ sech}^{2}\left[ \frac{J(q)m_{q}(t)+h_{q}(t)}{T}\right] ,
\end{equation}
 which in the linear limit can be written as

\begin{equation}
\label{lin dx/dt}
\frac{d\chi _{q}(t)}{dt}=\lambda ^{-1}\left[ \left( K(q)-1\right) \chi _{q}(t)+\frac{1}{T}\right] .
\end{equation}

Before we proceed with the solutions of these dynamical equations, we divide
the entire time zone in three different regimes : (I) \( 0<t<t_{0} \), where
\( h(t)=0 \) (II) \( t_{0}\leq t\leq t_{0}+\Delta t \), where \( h(t)=-h_{p} \)
and (III) \( t_{0}+\Delta t<t<\infty  \), where \( h(t)=0 \) again. We note
that (\ref{lin dm/dt}) can be readily solved separately for the three regions
as the boundary conditions are exactly known. In region I, \( dm/dt=0 \) and
the solution of the linearized (\ref{lin dm/dt}) becomes trivial. We, therefore,
use the solution of (\ref{dmq/dt}) in region I (\( m_{0}=\tanh \left( m_{0}T_{c}^{MF}/T\right)  \))
as the initial value of \( m \) for region II. Integrating (\ref{lin dm/dt})
in region II, we then get

\begin{equation}
\label{m(t) in II}
m(t)=\frac{h_{p}}{\Delta T}+\left( m_{0}-\frac{h_{p}}{\Delta T}\right) \exp \left[ b\Delta T\left( t-t_{0}\right) \right] ,
\end{equation}
 where \( b=1/\lambda T \) and \( \Delta T=T_{c}^{MF}-T \). It is to be noted
that in order to justify the validity of the linearization of (\ref{dmq/dt})
one must keep the factor inside the exponential of (\ref{m(t) in II}) small.
This restricts the linear theory to be valid at temperatures close to \( T_{c}^{MF} \)
and for small values of \( \Delta t \). Writing \( m_{w}\equiv m(t_{0}+\Delta t) \),
we get from (\ref{m(t) in II})

\begin{equation}
\label{mw}
m_{w}=\frac{h_{p}}{\Delta T}+\left( m_{0}-\frac{h_{p}}{\Delta T}\right) e^{b\Delta T\Delta t}.
\end{equation}
 It is to be noted here that in absence of fluctuations, the sign of \( m_{w}(h_{p},\Delta t) \)
solely decides which of the two final equilibrium states will be chosen by the
system after the withdrawal of the pulse. At \( t=t_{0}+\Delta t \), if \( m_{w}>0 \),
the system goes back to \( +m_{0} \) state and if \( m_{w}<0 \), magnetization-reversal
transition occurs and the system eventually chooses the \( -m_{0} \) state
(see figure 1). Thus setting \( m_{w}=0 \), we obtain the threshold value of
the pulse strength at the mean field phase boundary for this dynamic phase transition.
At any \( T \), combinations of \( h_{p} \) and \( \Delta t \) below the
phase boundary cannot induce the magnetization-reversal transition, while those
above it can induce the transition. From (\ref{mw}) therefore we can write
the equation of the mean field phase boundary for the magnetization-reversal
transition as

\begin{equation}
\label{phase-bound}
h_{p}^{c}(\Delta t,T)=\frac{\Delta Tm_{0}}{1-e^{-b\Delta T\Delta t}}.
\end{equation}
 Figure 2 shows phase boundaries at different \( T \) obtained from (\ref{phase-bound})
and compares those to the phase boundaries obtained from the numerical solution
of the full dynamical equation (\ref{dmq/dt}). The phase boundaries obtained
under linear approximation match quite well with those obtained numerically
for small values of \( \Delta t \) and at temperatures close to \( T_{c}^{MF} \),
which is the domain of validity of the linearized theory as discussed before.
In region III, we again have \( h(t)=0 \) and solution of (\ref{lin dm/dt})
leads to 

\begin{equation}
\label{m(t) in III}
m(t)=m_{w}\exp \left[ b\Delta T\left\{ t-\left( t_{0}+\Delta t\right) \right\} \right] .
\end{equation}
 We define the relaxation time \( \tau _{R}^{MF} \), measured from \( t=t_{0}+\Delta t \),
as the time required to reach the final equilibrium state characterized by magnetization
\( \pm m_{0} \) in region III (see figure 1). From (\ref{m(t) in III}) therefore
we can write

\begin{eqnarray}
\tau _{R}^{MF} & = & \frac{1}{b\Delta T}\ln \left( \frac{m_{0}}{\left| m_{w}\right| }\right) \nonumber \\
 & \sim  & -\left( \frac{T}{T_{c}^{MF}-T}\right) \ln \left| m_{w}\right| .\label{tau} 
\end{eqnarray}
A point to note is that \( m(t) \) in (\ref{m(t) in III}) grows exponentially
with \( t \) and therefore in order to confine ourselves to the linear regime
of \( m(t) \), \( m_{0} \) must be small (\( T \) close to \( T_{c}^{MF} \))
and \( t\leq \tau _{R}^{MF} \). The factor \( \left( T_{c}^{MF}-T\right) ^{-1} \)
gives the usual critical slowing down for the static transition at \( T=T_{c}^{MF} \).
However, even for \( T\ll T_{c}^{MF} \), \( \tau _{R}^{MF} \) diverges at
the magnetization-reversal phase boundary where \( m_{w} \) vanishes. Figure
3 shows the divergence of \( \tau ^{MF}_{R} \) against \( m_{w} \) as obtained
from the numerical solution of the full mean field equation of motion (\ref{dmq/dt})
and compares it with that obtained from (\ref{tau}).

Solution of \( \chi _{q}(t) \) is more difficult as all the boundary conditions
are not directly known. However, \( \chi _{q}(t) \) can be expressed in terms
of \( m(t) \) and the solution of the resulting equation will then have the
\( t \) dependence coming through \( m(t) \), which we have solved already.
Dividing (\ref{dxq/dt}) by (\ref{dmq/dt}) we get

\begin{equation}
\label{dx/dm}
\frac{d\chi _{q}(t)}{dm(t)}=\frac{-\chi _{q}(t)+\left( \frac{J(q)\chi _{q}(t)+1}{T}\right) \textrm{ sech}^{2}\left[ \frac{J(q)m_{q}(t)+h_{q}(t)}{T}\right] }{-m_{q}(t)+\tanh \left( \frac{J(q)m_{q}(t)+h_{q}(t)}{T}\right) },
\end{equation}
 which can be rewritten in the linear limit as

\begin{equation}
\label{lin dx/dm}
\frac{d\chi _{q}}{\chi _{q}+\Gamma }=a_{q}\frac{dm}{m+h(t)/\Delta t},
\end{equation}
 where \( \Gamma =1/T\left( K(q)-1\right)  \) and \( a_{q}=\left( K(q)-1\right) /\left( K(0)-1\right) \simeq 1-q^{2}/\Delta T \)
for small \( q \).

In region II, solution of (\ref{lin dx/dm}) can be written as

\begin{equation}
\label{x(t) in II}
\chi _{q}(t)=-\Gamma +\left( \chi _{q}^{s}+\Gamma \right) \left[ \frac{m(t)-h_{p}/\Delta T}{m_{0}-h_{p}/\Delta T}\right] ^{a_{q}},
\end{equation}
 where \( \chi _{q}^{s} \) is the equilibrium value of susceptibility in region
I. Solving (\ref{lin dx/dm}) in region III with the initial boundary condition
\( m\left( t_{0}+\Delta t\right) =m_{w} \), we get

\begin{eqnarray}
\chi _{q}(t) & = & -\Gamma +\left( \chi _{q}\left( t_{0}+\Delta t\right) +\Gamma \right) \left( \frac{m(t)}{m_{w}}\right) ^{a_{q}}\nonumber \\
 & = & -\Gamma +\left( \chi _{q}^{s}+\Gamma \right) \left( \frac{m(t)}{m_{w}}\right) ^{a_{q}}e^{b\Delta T\Delta ta_{q}},\label{x(t) in III} 
\end{eqnarray}
 where use has been made of (\ref{x(t) in II}) and (\ref{mw}). The dominating
\( q \) dependence in \( \chi _{q}(t) \) is coming from \( \left( 1/m_{w}\right) ^{a_{q}} \)
when \( m_{w}\rightarrow 0 \) as one approaches the phase boundary. The singular
part of the dynamic susceptibility can then be written as 

\begin{equation}
\label{x(q) with xi}
\chi _{q}(t)=\left( \chi _{q}^{s}+\Gamma \right) \exp \left[ -q^{2}\left( \xi ^{MF}\right) ^{2}\right] ,
\end{equation}
 where for small values of \( m_{w} \) the correlation length \( \xi ^{MF} \)
is given by \cite{smc99}

\begin{equation}
\label{xi}
\xi ^{MF}\equiv \xi ^{MF}\left( m_{w}\right) =\left[ \frac{T_{c}}{\Delta T}\ln \left( \frac{1}{\left| m_{w}\right| }\right) \right] ^{\frac{1}{2}}.
\end{equation}
 Thus the length scale also diverges at the magnetization-reversal phase boundary
and this can be demonstrated even using the linearized mean field equation of
motion. Equations (\ref{tau}) and (\ref{xi}) can now be used to establish
the following relation between the diverging time and length scales :

\begin{equation}
\label{tau-xi}
\tau _{R}^{MF}\sim \frac{T}{T_{c}}\left( \xi ^{MF}\right) ^{2},
\end{equation}
 which leads to a dynamical critical exponent \( z=2. \) It may be noted that
these divergences in \( \tau _{R}^{MF} \) and \( \xi ^{MF} \) are shown to
occur for any \( T<T_{C}^{MF} \), and these dynamic relaxation time and correlation
length defined for the magnetization-reversal transition exist only for \( T<T_{c}^{MF} \).

It may further be noted from (\ref{x(q) with xi}) that \( \chi _{q}(t)\rightarrow 0 \)
as \( \xi ^{MF}\rightarrow \infty  \), thereby producing a minimum of \( \chi _{q} \)
at the phase boundary. The absence of any divergence in the susceptibility is
due to the fact that at \( t=t_{0}+\Delta t \), there remains no contribution
of \( m_{w} \) in \( \chi _{q}(t) \) as is evident from (\ref{x(t) in III}).
However, numerical solution of (\ref{dx/dm}) for \( q=0 \) mode shows a clear
singularity in the homogeneous susceptibility \( \chi _{0} \) at the magnetization-reversal
phase boundary (\( m_{w}=0 \)), as depicted in figure 4. One can also have
a numerical estimate of \( \xi ^{MF} \) by solving (\ref{dx/dm}) for different
values of \( q \). Figure 5 shows plots of \( \chi _{q}(t) \) against \( m_{w} \)
for different values of \( q \). The inset of figure 5 shows the variation
of \( \left( \xi ^{MF}\right) ^{-2} \) against \( \left( \ln \left| m_{w}\right| \right) ^{-1} \),
where \( \xi ^{MF} \) was obtained by fitting the data of figure 5 with straight
lines. It is clearly seen from the inset that for small values of \( m_{w} \)
the linear approximation agrees quite well with the numerical results.

\section{Monte Carlo Study}

We now study the transition using Monte Carlo simulation with single spin-flip
Glauber dynamics \cite{binder84}. Working at a temperature below the static
critical temperature (\( T^{0}_{c}\simeq 2.27 \) and \( 4.51 \) \cite{blh95}
in units of the nearest neighbour interaction strength \( J \) for square and
simple cubic lattices respectively), the system is prepared by evolving the
initial state (say with all spins up) under Glauber dynamics for the temperature
\( T \). The evolution time \( t_{0} \) is usually taken to be sufficiently
larger than the static relaxation time at \( T \) to ensure that the system
reaches an equilibrium state with magnetization \( m_{0} \) before the external
magnetic field is applied at time \( t=t_{0} \). The magnetization \( m(t) \)
starts decreasing from its initial value \( m_{0} \) due to the effect of the
competing field during the period \( t_{0}\leq t\leq t_{0}+\Delta t \), and
it assumes the value \( m_{w} \) at \( t=t_{0}+\Delta t \). Due to presence
of fluctuations, \( m_{w}<0 \) does not necessarily lead to a magnetization-reversal
whereas even for \( m_{w}>0 \) fluctuations can give rise to a magnetization-reversal.
This is in contrast with the mean field case, where due to the absence of any
fluctuation the sign of \( m_{w} \) solely determines the final state. In the
MC study, however, on an average the final state is determined by the sign of
\( m_{w} \) (see figure 1). The magnetization-reversal transition phase boundary
therefore again corresponds to \( m_{w}=0 \).

Figure 6 shows phase boundaries at different temperatures for square and simple
cubic lattices. The data points for \( d=2 \) are averaged over 500 different
Monte Carlo runs (MCR) and those for \( d=3 \) are averaged over 150 MCR. A
qualitative difference between the MF and the MC phase boundaries may be noted
here. In the former, even for \( \Delta t\rightarrow \infty  \), due to the
absence of fluctuations, \( h_{p} \) must be greater than the non-zero coercive
field to bring about the transition and therefore the phase boundaries flatten
for larger values of \( \Delta t \). However, in real systems fluctuations
are present and even an infinitesimal strength of the pulse, if applied for
very long time, can bring about the transition. This is evident from the asymptotic
nature of the phase boundaries for large values of \( \Delta t \).

It is instructive to look at the classical theory of nucleation to understand
the nature of the MC phase diagram of the magnetization-reversal transition.
A typical configuration of a ferromagnet, below its static critical temperature
\( T_{c}^{0} \), consists of droplets or domains of spins oriented in the same
direction, in a sea of oppositely oriented spins. According to CNT, the equilibrium
number of droplets consisting of \( s \) spins is given by \( n_{s}=N\exp \left( -\epsilon _{s}/T\right)  \),
where \( \epsilon _{s} \) is the free energy of formation of a droplet containing
\( s \) spins and \( N \) is a normalization constant. In presence of a negative
external magnetic field \( h \), the free energy can be written as \( \epsilon _{s}=-2hs+\sigma s^{(d-1)/d} \),
where the shape of the droplet is assumed to be spherical and \( \sigma (T) \)
is the temperature dependent surface tension. Droplets of size greater than
a critical value \( s_{c} \) are favoured to grow, where \( s_{c}=\left[ \sigma (d-1)/(2d\left| h\right| )\right] ^{d} \)
is obtained by maximizing \( \epsilon _{s} \). The number of supercritical
droplets is therefore given by \( n_{s_{c}}=N\exp \left[ -\Lambda _{d}\sigma ^{d}\left| h\right| ^{1-d}/T\right]  \),
where \( \Lambda _{d} \) is a constant depending on dimension only. In the
SD regime, where a single supercritical droplet grows to engulf the whole system,
the nucleation time is inversely proportional to the nucleation rate \( I \).
According to the Becker-D\( \ddot{\textrm{o}} \)ring theory, \( I \) is proportional
to \( n_{s_{c}} \) and therefore one can write
\[
\tau ^{SD}_{N}\propto I^{-1}\propto \exp \left[ \frac{\Lambda _{d}\sigma ^{d}}{T\left| h\right| ^{d-1}}\right] .\]
 However, in the MD regime the nucleation mechanism is different and in this
regime many supercritical droplets grow simultaneously and eventually coalesce
to create a system spanning droplet. The radius \( s_{c}^{1/d} \) of a supercritical
droplet grows linearly with time \( t \) and thus \( s_{c}\propto t^{d} \).
For a steady rate of nucleation, the rate of change of magnetization is \( It^{d} \).
For a finite change \( \Delta m \) of the magnetization during the nucleation
time \( \tau _{N}^{MD} \), one can write

\[
\Delta m\propto \int _{0}^{\tau ^{MD}_{N}}It^{d}dt=I\left( \tau _{N}^{MD}\right) ^{d+1}.\]
 Therefore, in the MD regime one can write \cite{rtms94}\cite{as98} 
\[
\tau _{N}^{MD}\propto I^{-1/(d+1)}\propto \exp \left[ \frac{\Lambda _{d}\sigma ^{d}}{T(d+1)\left| h\right| ^{d-1}}\right] .\]

During the time \( t_{0}\leq t\leq t_{0}+\Delta t \), when the external field
remains `on', the only relevant time scale in the system is the nucleation time.
The magnetization reversal phase boundary gives the threshold value \( h_{p}^{c} \)
of the pulse strength which, within time \( \Delta t \), brings the system
from an equilibrium state with magnetization \( +m_{0} \) to a non-equilibrium
state with magnetization \( m_{w}=0_{-} \), so that eventually the system evolves
to the equilibrium state with magnetization \( -m_{0} \). The field driven
nucleation mechanism takes place for \( t_{0}\leq t\leq t_{0}+\Delta t \) and
therefore equating the above nucleation times with \( \Delta t \), one gets
the for the magnetization-reversal phase boundary
\begin{equation}
\label{logdt-hp}
\begin{array}{cccccl}
\ln \left( \Delta t\right)  & = & c_{1} & + & C\left[ h_{p}^{c}\right] ^{1-d}, & \textrm{in the SD regime}\\
 & = & c_{2} & + & C\left[ h_{p}^{c}\right] ^{1-d}/(d+1), & \textrm{in the MD regime}
\end{array}
\end{equation}

\noindent where \( C=\Lambda _{d}\sigma ^{d}/T \) and \( c_{1} \), \( c_{2} \)
are constants. Therefore a plot of \( \ln (\Delta t) \) against \( \left[ h_{p}^{c}\right] ^{d-1} \)
would show two different slopes corresponding to the two regimes \cite{mc00}.
Figure 7 shows these plots and it indeed have two distinct slopes for both \( d=2 \)
(figure 7(a)) and \( d=3 \) (figure 7(c)) at sufficiently high temperatures,
where both the regimes are present. The ratio \( R \) of the slopes corresponding
to the two regimes has got values close to \( 3 \) for \( d=2 \) and close
to \( 4 \) for \( d=3 \), as suggested by (\ref{logdt-hp}). The value of
\( h_{DSP} \) is obtained from the point of intersection of the straight lines
fitted to the two regimes. At lower temperatures, however, the MD region is
absent and the phase diagram here is marked by a single slope as shown in figures
7(b) and 7(d).

Once the pulse is withdrawn, the system relaxes to one of the two equilibrium
states. The closer one leaves the system to the phase boundary (\( m_{w}\rightarrow 0 \)),
larger is the relaxation time \( \tau _{R} \). However, unlike the mean field
case, the MC relaxation time falls off exponentially with \( \left| m_{w}\right|  \)
away from the phase boundary. Figure 8 shows the growth of \( \tau _{R} \)
as \( m_{w}\rightarrow 0 \) at a particular \( T \) and for a particular \( \Delta t \).
Typical number of MCR used to obtain the data is \( 400 \) for \( L=40 \)
and \( 25 \) for \( L=400 \). The best fitted curve through the data points
shows the relaxation behaviour as follows :
\begin{equation}
\label{MCtau}
\tau _{R}\sim \kappa (T,L)e^{-\mu (T)\left| m_{w}\right| },
\end{equation}
 where \( \kappa (T,L) \) is a constant depending on temperature and system
size and \( \mu (T) \) is a constant depending on temperature only. It may
be noted from (\ref{MCtau}) that \( \tau \rightarrow \kappa (T,L) \) as \( m_{w}\rightarrow 0 \).
Therefore the true divergence at the phase boundary (where \( m_{w}=0 \)) of
the relaxation time depends on the nature of \( \kappa (T,L) \). The inset
of figure 8 shows the sharp growth of \( \kappa (T,L) \) with the system size.
The relaxation time \( \tau _{R} \) therefore diverges in the thermodynamic
limit (\( L\rightarrow \infty  \)) through the constant \( \kappa  \). It
may be noted that this divergence of \( \tau _{R} \) at the dynamic magnetization-reversal
phase boundary occurs even at temperatures far below the static critical temperature
\( T_{c}^{0} \). 

According to CNT, \( s_{c}\propto \left| h_{p}\right| ^{-d} \) and therefore
at any fixed \( T \), stronger fields will allow many critical droplets to
form and hence the system goes over to the MD regime. On the other hand, a weaker
field rules out the possibility of more than one critical droplet and therefore
the system goes over to the SD regime. Figure 9 shows snapshots of the spin
configurations at different times in both SD and MD regimes. The snapshots at
\( t=t_{0}+\Delta t \) corresponds to \( m_{w}\sim O\left( 10^{-2}\right)  \).
\( h_{p}>h_{DSP} \) in figure 9(a) and a single large droplet is formed whereas
\( h_{p}<h_{DSP} \) in (b) and many droplets are seen to be formed. It may
be noticed from figure 9 that the boundaries of the droplets are flat with very
few kinks on it at \( t=t_{0}+\Delta t \). The probability of growth of a droplet
along a flat boundary is very small (only \( 25\% \) in case of a square lattice)
and hence domain wall movement practically stops immediately after the withdrawal
of the field. This restricts further nucleation. It is then left to very large
fluctuations to resume the domain wall movement and long time is required for
the system to come out of the metastable state and subsequently reach the final
equilibrium state. Thus the effect of the pulse is to initiate the nucleation
process and the threshold value of the pulse strength is such that within the
pulse duration it renders the system with droplets almost without any kink in
it. This observation justifies the sharp growth of the relaxation time at the
phase boundary. 

The growth of a length scale at the transition phase boundary can be qualitatively
shown from the distribution of domains of reversed spins. We define a pseudo-correlation
length \( \widetilde{\xi } \) as 

\begin{equation}
\label{xitilde}
\widetilde{\xi }^{2}=\frac{\sum _{s}R_{s}^{2}s^{2}n_{s}}{\sum _{s}s^{2}n_{s}},
\end{equation}
 where the radius of gyration \( R_{s} \) is defined as \( R_{s}^{2}=\sum _{i=1}^{s}\left| r_{i}-r_{0}\right| ^{2}/s \),
\( r_{i} \) denoting the position vector of the \( i \)th spin of the domain
and \( r_{0}=\sum _{i=1}^{s}r_{i}/s \) being the centre of mass of the domain.
As the transition phase boundary is approached, \( \widetilde{\xi } \) is observed
to grow with the system size as shown in figure 10. Typical number of MCR used
for obtaining the data is \( 10 \) for \( L=1000 \) and \( 2000 \) for \( L=50 \).
This indicates the divergence of a length scale at the phase boundary in the
thermodynamic limit. It should be noted, however, that \( \widetilde{\xi } \)
is not exactly the correlation length of the system \cite{robin85}. An estimate
for the power law growth of the actual correlation length \( \xi  \), as the
phase boundary is approached in the MD region, will be obtained from the finite
size scaling study discussed later in this section.

The order of the magnetization-reversal transition changes with temperature
and with \( \Delta t \) even along the same phase boundary. The transition
is discontinuous all along the low \( T \) phase boundary, whereas at higher
values of \( T \) the nature of the transition changes from a continuous to
a discontinuous one as one moves towards higher values of \( \Delta t \). For
\( h_{p}^{c}(T)\ll h_{DSP}(T) \), the system is brought to the SD regime where
the order of the transition is observed to be discontinuous. On the other hand
continuous transition is observed for \( h_{p}^{c}(T)\gg h_{DSP}(T) \) when
the system goes over to the MD regime. One can look at the probability distribution
\( P(m_{w}) \) of \( m_{w} \) to determine the order of the phase transition.
Figure 11 shows the variation of \( P(m_{w}) \) as the phase boundary corresponding
to a particular temperature is crossed at two different positions (different
\( \Delta t \)). The data are averaged over \( 500 \) MCR. The existence of
a single peak in (a), which shifts its position continuously from \( +1 \)
to \( -1 \) as the phase boundary is crossed, indicates the continuous nature
of the transition. In (b), however, two peaks of comparable strength at positions
close to \( \pm m_{0} \) exist simultaneously. This shows that the system can
simultaneously reside in both the phases which is a sure indication for a discontinuous
phase transition. On phase boundaries corresponding to higher temperatures the
crossover from the discontinuous transition to a continuous one is not very
sharp and there exists a region around \( h_{p}^{c}=h_{DSP} \) on the phase
boundary, over which the nature of the transition cannot be determined with
certainty. This is evident from figure 7, where the data points near the tricritical
point do not fit to the slope of either of the straight lines corresponding
to the two different regimes.

In the region where the transition is continuous in nature one can expect scaling
arguments to hold. We assume power law behaviour in this regime both for \( m_{w} \)
\begin{equation}
\label{powerlaw mw}
m_{w}\sim \left| h_{p}-h_{p}^{c}\left( \Delta t,T\right) \right| ^{\beta }
\end{equation}
 and for the correlation length 
\begin{equation}
\label{powerlaw xi}
\xi \sim \left| h_{p}-h_{p}^{c}\left( \Delta t,T\right) \right| ^{-\nu }.
\end{equation}
For a finite size system, \( h_{p}^{c} \) is a function of the system size
\( L \). Assuming that at the phase boundary \( \xi  \) can at the most reach
a value equal to \( L \), one can write the finite size scaling form of \( m_{w} \)
as \cite{barber83} :

\begin{equation}
\label{fss}
m_{w}\sim L^{-\beta /\nu }f\left[ \left( h_{p}-h^{c}_{p}\left( \Delta t,T,L\right) \right) L^{1/\nu }\right] ,
\end{equation}
 where \( f(x)\sim x^{\beta /\nu } \) as \( x\rightarrow \infty  \). A plot
of \( m_{w}/L^{-\beta /\nu } \) against \( \left( h_{p}-h^{c}_{p}\left( \Delta t,T,L\right) \right) L^{1/\nu } \)
shows a nice collapse of the data corresponding to \( L=50 \), \( 100 \),
\( 200 \), \( 400 \) and \( 800 \) for \( d=2 \) and \( L=10 \), \( 20 \),
\( 40 \), \( 80 \) and \( 120 \) for \( d=3 \) as shown in figure 12. Typical
number of MCR used to obtain the data is \( 5120 \) for \( L=50 \) in \( d=2 \)
and \( 10000 \) for \( L=10 \) in \( d=3 \). The values of the critical exponents
obtained from the data collapse are \( \beta =0.85\pm 0.05 \) and \( \nu =1.5\pm 0.5 \)
in \( d=3 \) and \( \beta =1.00\pm 0.05 \) and \( \nu =2.0\pm 0.5 \) in \( d=2 \),
where \( h_{p}^{c}\left( \Delta t,T\right)  \) was obtained with an accuracy
\( O\left( 10^{-3}\right)  \). All attempts to fit similar data to the above
finite size scaling form obtained in the SD regime failed.

The accuracy with which \( h_{p}^{c}\left( \Delta t,T\right)  \) is measured,
is very crucial for obtaining the critical exponents through finite size scaling.
The cumulant method introduced by Binder et al. \cite{binder88} is one of the
reliable methods which can be employed to obtain the value of \( h_{p}^{c} \).
The fourth order cumulant is defined as

\begin{equation}
\label{bincum}
g(L)=\frac{1}{2}\left[ 3-\frac{\left\langle m^{4}_{w}\right\rangle }{\left\langle m^{2}_{w}\right\rangle ^{2}}\right] ,
\end{equation}
 where \( \left\langle m_{w}^{n}\right\rangle =\int m_{w}^{n}P(m_{w})dm_{w} \).
The quantity \( g(L) \) is dimensionless and is equal to unity for \( \left| m_{w}\right| \gg 0 \),
while \( g(L)\rightarrow 0 \) for \( m_{w}\rightarrow 0 \), assuming a Gaussian
distribution of \( m_{w} \) around \( 0 \) on the phase boundary. Figure 13
shows a plot of \( g(L) \) against \( h_{p} \) at a fixed \( \Delta t \)
and \( T \) and the value of the pulse strength corresponding to the point
of intersection of the different curves gives \( h_{p}^{c}(\Delta t,T) \);
assuming \( g\equiv g\left[ L/\left| h_{p}-h_{p}^{c}\right| ^{-\nu }\right]  \).
Typical number of MCR used to obtain the data is \( 50000 \) for \( L=50 \)
and \( 2500 \) for \( L=800 \). It is to be noted that none of the curves
touch the abscissa which corresponds to \( m_{w}=0 \) which is numerically
unattainable. The closer one gets to \( m_{w}=0 \) better the accuracy in the
measurement of \( h_{p}^{c} \). In principle the minima of \( g(L) \) correponding
to different \( L \) should occur at the same position (at \( h_{p}=h_{p}^{c} \)).
The shift in the position of the minima of \( g(L) \) in figure 13 is caused
by the presence of large fluctuations in measuring higher moments of \( m_{w} \).
However, this estimate of \( h_{p}^{c} \), when used in the scaling fit of
(\ref{fss}), did not significantly improve the estimates of the critical exponents
\( \beta  \) and \( \nu  \).

\section{Summary and Conclusions}

In this paper we have discussed in detail almost all the studies that have been
made so far on the dynamic magnetization-reversal transition in the Ising model
under finite duration external magnetic field competing with the existing order
for \( T<T_{c}^{0} \). Any combination of the pulse strength and duration above
the phase boundary in the \( h_{p}-\Delta t \) plane leads to the transition
from one ordered phase to the equivalent other. We solved numerically the mean
field equation of motion for the magnetization to obtain the MF phase boundary
where the susceptibility and the relaxation time were observed to diverge. The
divergence of both the time (\( \tau ^{MF}_{R} \)) and the length scale (\( \xi ^{MF} \))
at the MF phase boundary was observed even from the analytic solution of the
MF equations of motion under a linear approximation. Under this approximation,
the dynamical critical exponent was found to have a value \( 2 \) : \( \tau _{R}^{MF}\sim \left( \xi ^{MF}\right) ^{2}\sim -\ln \left| m_{w}\right|  \),
where \( m_{w}\left( h_{p},\Delta t,T\right) =0 \) gives the phase boundary.
The same transition has been studied using Monte Carlo simulations in both two
and three dimensions. The obtained phase diagram is fully consistent with the
classical nucleation theory. The nucleation process is initiated by the external
magnetic field and depending on the strength of the field the system nucleates
either through the growth of a single droplet or through the growth and subsequent
coalescence of many droplets. For \( h_{p}>h_{DSP} \) the system belongs to
the multi-droplet regime and the transition is continuous in nature; whereas
for \( h_{p}<h_{DSP} \) the system goes over to the single-droplet regime where
transition is discontinuous. Expecting power law behaviour for both \( m_{w} \)
and \( \xi  \) in multi-droplet regime, the finite size scaling fits give the
estimates of the critical exponents \( \beta  \) and \( \nu  \) for both \( d=2 \)
and \( 3 \). Unlike in the MF case, where the relaxation time \( \tau _{R}^{MF} \)
shows a logarithmic divergence, \( \tau _{R} \) in MC studies falls off exponentially
away from \( m_{w}=0 \) and the divergence in \( \tau _{R} \) comes through
the growth of the prefactor \( \kappa  \) in (\ref{MCtau}) with the system
size.

The symmetry breaking transition of the dynamic hysteresis in pure Ising systems
under oscillating external fields \cite{ca99}\cite{ac95}\cite{srn98.b}, where
the \( m-h \) loop becomes asymmetric due to the fact that the magnetization
\( m(t) \) fails to follow even the phase or sign of the rapidly changing field
\( h(t) \), leads to a dynamic transition. This dynamic transition has been
studied employing finite size scaling theory \cite{srn98.b}\cite{srn99} and
the estimates of the critical exponents seem to be consistent with the static
Ising universality class \cite{rkwns00}. Although this transition as well as
the one discussed in this paper occur due to the failure of the system to get
out of the `free energy well' corresponding to the existing order because of
the lack of proper combination of the pulse strength and duration, they belong
to different universality classes.

{\par\raggedright \textbf{\large Acknowledgements}\large \par}

\noindent We are grateful to M. Acharyya, D. Chowdhury, C. Dasgupta, B. Duenweg,
D. Stauffer and R. B. Stinchcombe for their useful comments and suggestions.

\newpage
\noindent \textbf{\large Figure Captions}{\large \par}

\noindent \textbf{Figure 1.} Typical time variation of the response magnetizations
\( m(t) \) for two different field pulses \( h(t) \) with same \( \Delta t \)
and \( T \) are shown. The quantities of interest to characterize the response
magnetizations for both the pulses are indicated.

\noindent \textbf{Figure 2.} MF phase boundaries for three different temperatures.
The solid line is obtained from numerical solution of (\ref{dmq/dt}) and the
dotted lines give the corresponding analytical estimates in the linear limit.

\noindent \textbf{Figure 3.} Logarithmic divergence of \( \tau _{R}^{MF} \)
across the phase boundary for \( T/T_{c}=0.9 \). The data points shown by circles
are obtained from the solution of (\ref{dmq/dt}) and the solid line corresponds
to the solution of the linearized MF equation.

\noindent \textbf{Figure 4.} Divergence of \( \chi _{q=0} \) across the phase
boundary obtained from the numerical solution of (\ref{dx/dm}).

\noindent \textbf{Figure 5.} Plot of \( \chi _{q} \) against \( m_{w} \) for
different values of \( q \). The inset shows the linear variation of \( \left( \xi ^{MF}\right) ^{-2} \)
against \( \left[ \ln \left| m_{w}\right| \right] ^{-1} \). The data points
for \( \xi ^{MF} \) in the inset are obtained from the slope of the best fitted
straight lines through a plot of \( \ln \chi _{q} \) against \( q^{2} \) for
different values of \( m_{w} \).

\noindent \textbf{Figure 6.} Phase boundaries obtained from the MC study for
(a) square lattice with \( L=100 \) and (b) simple cubic lattice with \( L=50 \).

\noindent \textbf{Figure 7.} Plot of \( \ln \Delta t \) against \( \left( h_{p}\right) ^{1-d} \)
along the MC phase boundary. (a) \( T/T_{c}=0.31 \) and (b) \( T/T_{c}=0.09 \)
for square lattice and (c) \( T/T_{c}=0.67 \) and (d) \( T/T_{c}=0.11 \) for
simple cubic lattice. The slope ratio \( R\simeq 3.27 \) in (a) and \( \simeq 3.97 \)
in (c).

\noindent \textbf{Figure 8.} MC results for the divergence of \( \tau _{R} \)
for \( L=40 \), \( 50 \), \( 100 \), \( 200 \) and \( 400 \). The best
fitted straight lines are guide to the eye. The inset shows the variation with
\( L \) of the peak height \( \kappa  \) in the prefactor of \( \tau _{R} \)
in (\ref{MCtau}).

\noindent \textbf{Figure 9.} Snapshots of spin configurations in a \( 100\times 100 \)
square lattice at different stages (\( t=t_{0} \), \( t_{1} \) and \( t_{0}+\Delta t \)
) of nucleation, where \( t_{0}<t_{1}<t_{0}+\Delta t \). The dots correspond
to \( +1 \) spin state. (a) \( h_{p}=0.55, \) \( \Delta t=300 \) at \( T/T_{c}=0.44 \)
(SD regime) and (b) \( h_{p}=0.52 \), \( \Delta t=9 \) at \( T/T_{c}=0.88 \)
(MD regime).

\noindent \textbf{Figure 10.} Variation of \( \widetilde{\xi } \) with \( L \)
for \( L=50 \), \( 100 \), \( 200 \), \( 400 \), \( 800 \) and \( 1000 \)
for MC study on a square lattice.

\noindent \textbf{Figure 11.} Plot of \( P(m_{w}) \) against \( m_{w} \) as
one crosses the phase boundary for the MC study on a \( 100\times 100 \) square
lattice in (a) MD regime and (b) SD regime.

\noindent \textbf{Figure 12.} Finite size scaling fits : (a) for \( d=2 \)
at \( T/T_{c}=0.88 \) and (b) for \( d=3 \) at \( T/T_{c}=0.67 \).

\noindent \textbf{Figure 13.} Plot of \( g(L) \) against \( L \) for \( L=50 \),
\( 100 \), \( 200 \), \( 400 \) and \( 800 \) in the MD regime for MC study
on a \( 100\times 100 \) square lattice at \( T=2.0 \) for \( \Delta t=5 \).
\vspace{0.5001cm}\newpage

\vspace{0.5001cm}
{\par\centering \resizebox*{0.7\textwidth}{0.7\textheight}{\includegraphics{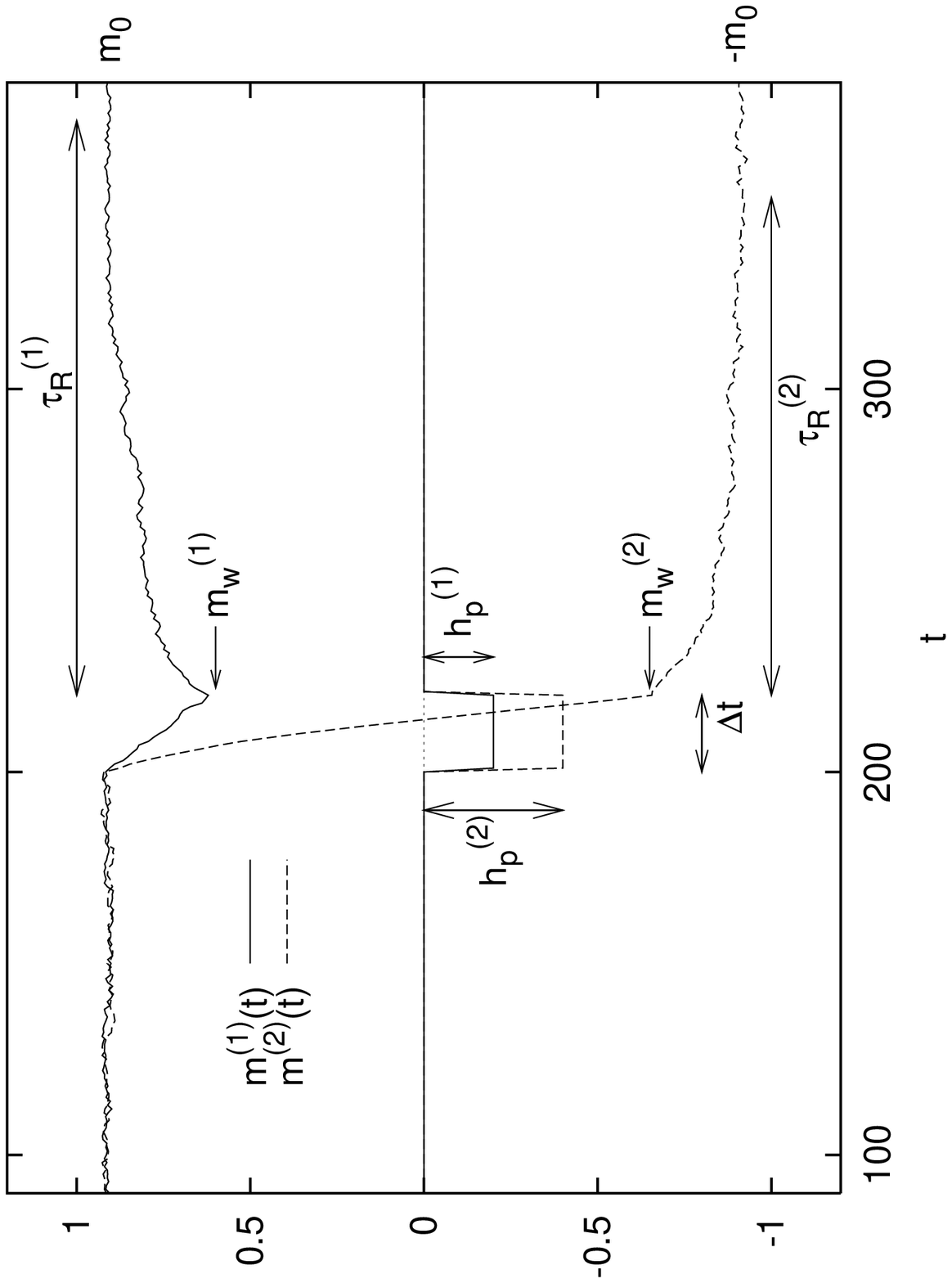}} \par}
\vspace{0.5001cm}

{\par\noindent \centering \textbf{Figure 1.}\par}
\vspace{0.5001cm}\newpage

\vspace{0.5001cm}
{\par\centering \resizebox*{0.7\textwidth}{0.7\textheight}{\includegraphics{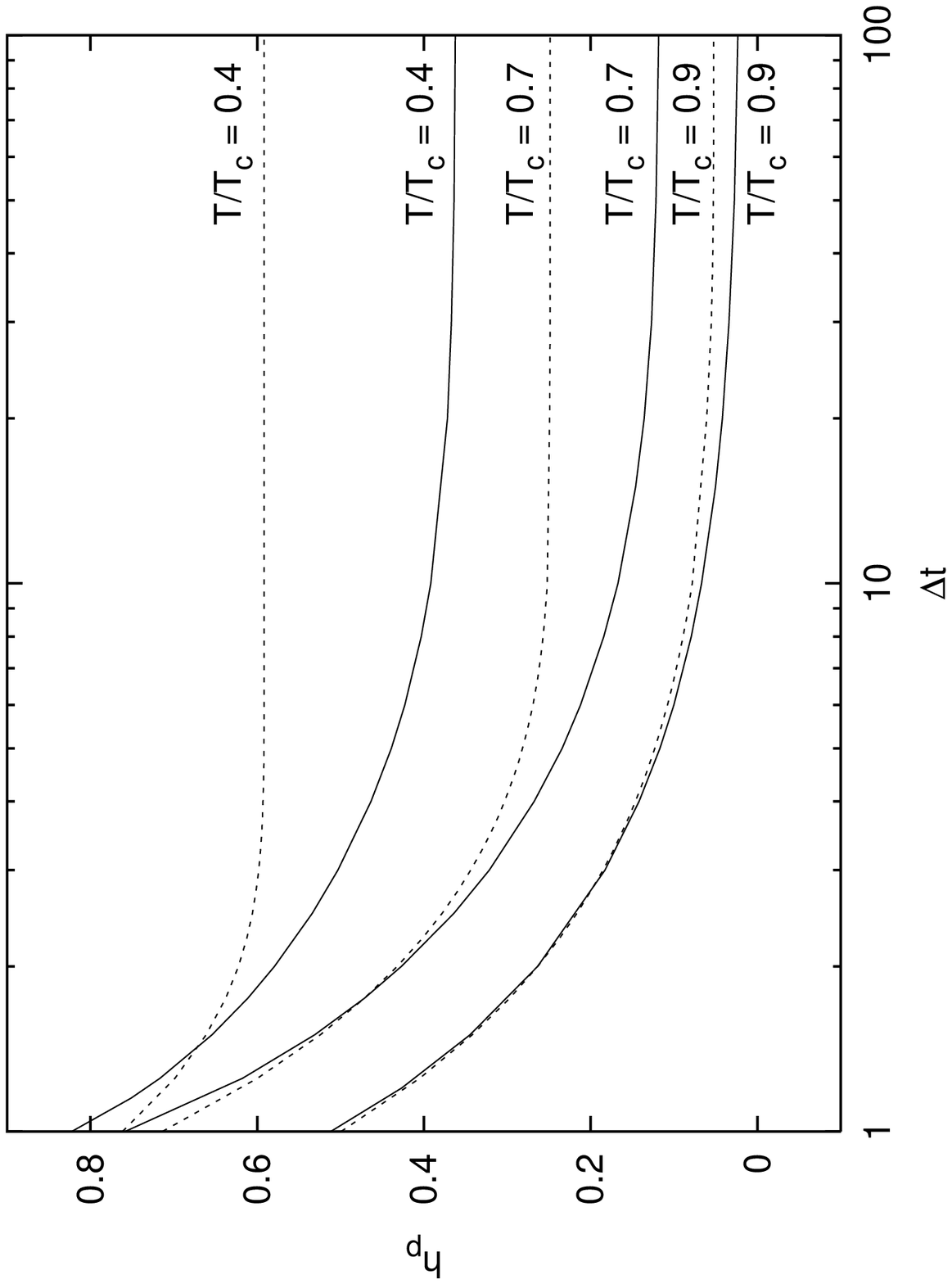}} \par}
\vspace{0.5001cm}

{\par\noindent \centering \textbf{Figure 2.}\par}
\vspace{0.5001cm}\newpage

\vspace{0.5001cm}
{\par\centering \resizebox*{0.7\textwidth}{0.7\textheight}{\includegraphics{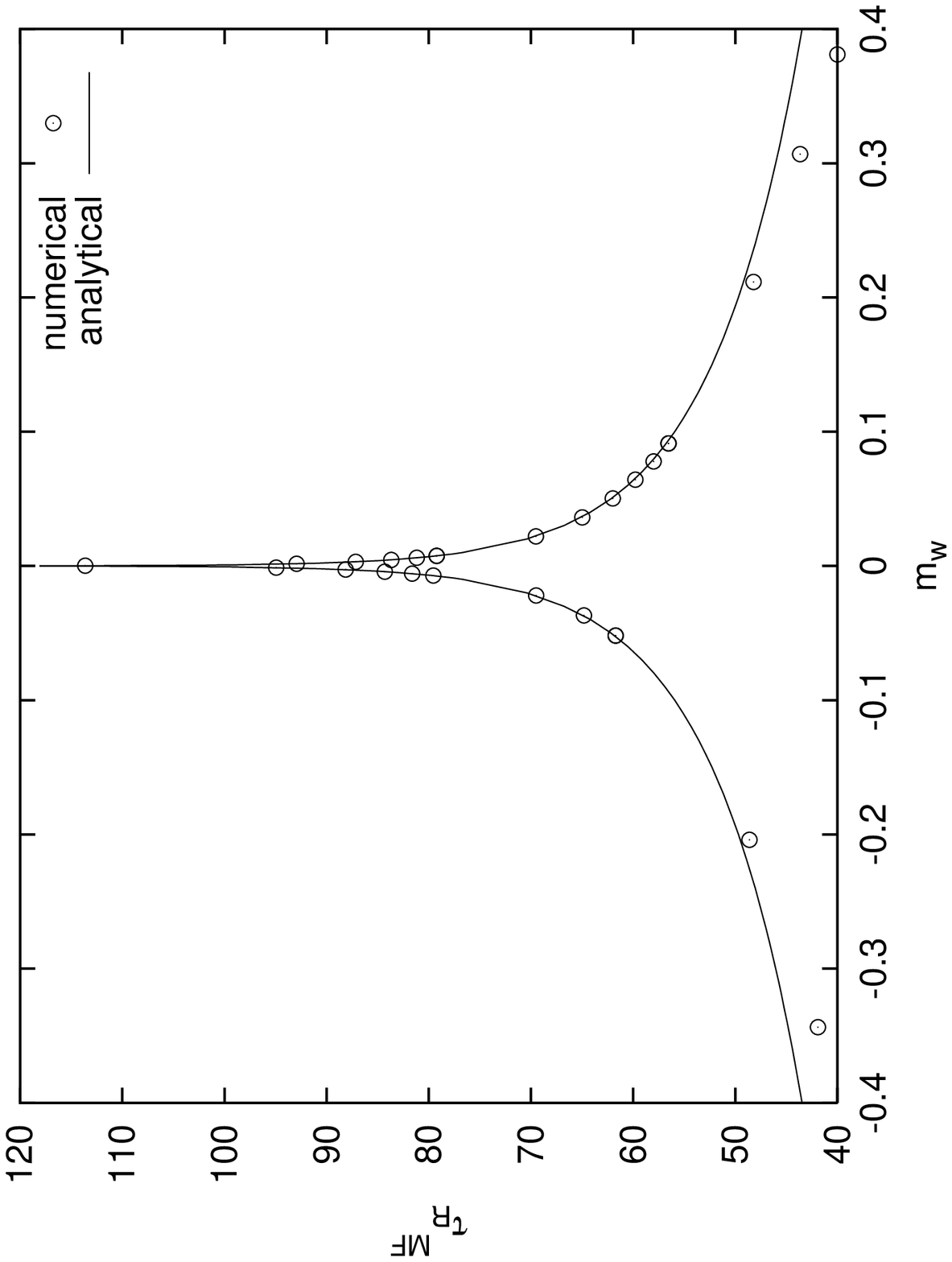}} \par}
\vspace{0.5001cm}

{\par\noindent \centering \textbf{Figure 3.}\par}
\vspace{0.5001cm}\newpage

\vspace{0.5001cm}
{\par\centering \resizebox*{0.7\textwidth}{0.7\textheight}{\includegraphics{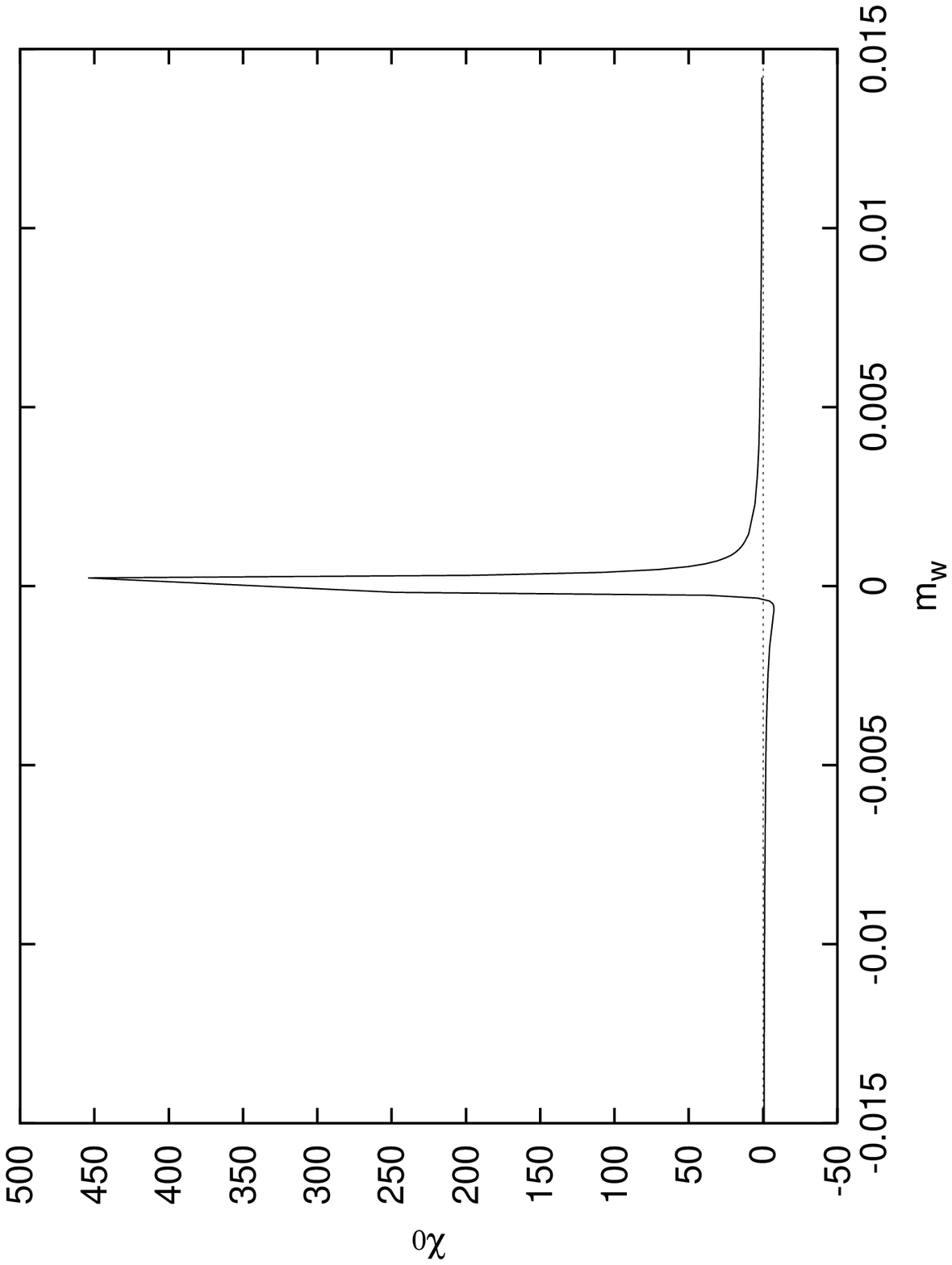}} \par}
\vspace{0.5001cm}

{\par\noindent \centering \textbf{Figure 4.}\par}
\vspace{0.5001cm}\newpage

\vspace{0.5001cm}
{\par\centering \resizebox*{0.7\textwidth}{0.7\textheight}{\includegraphics{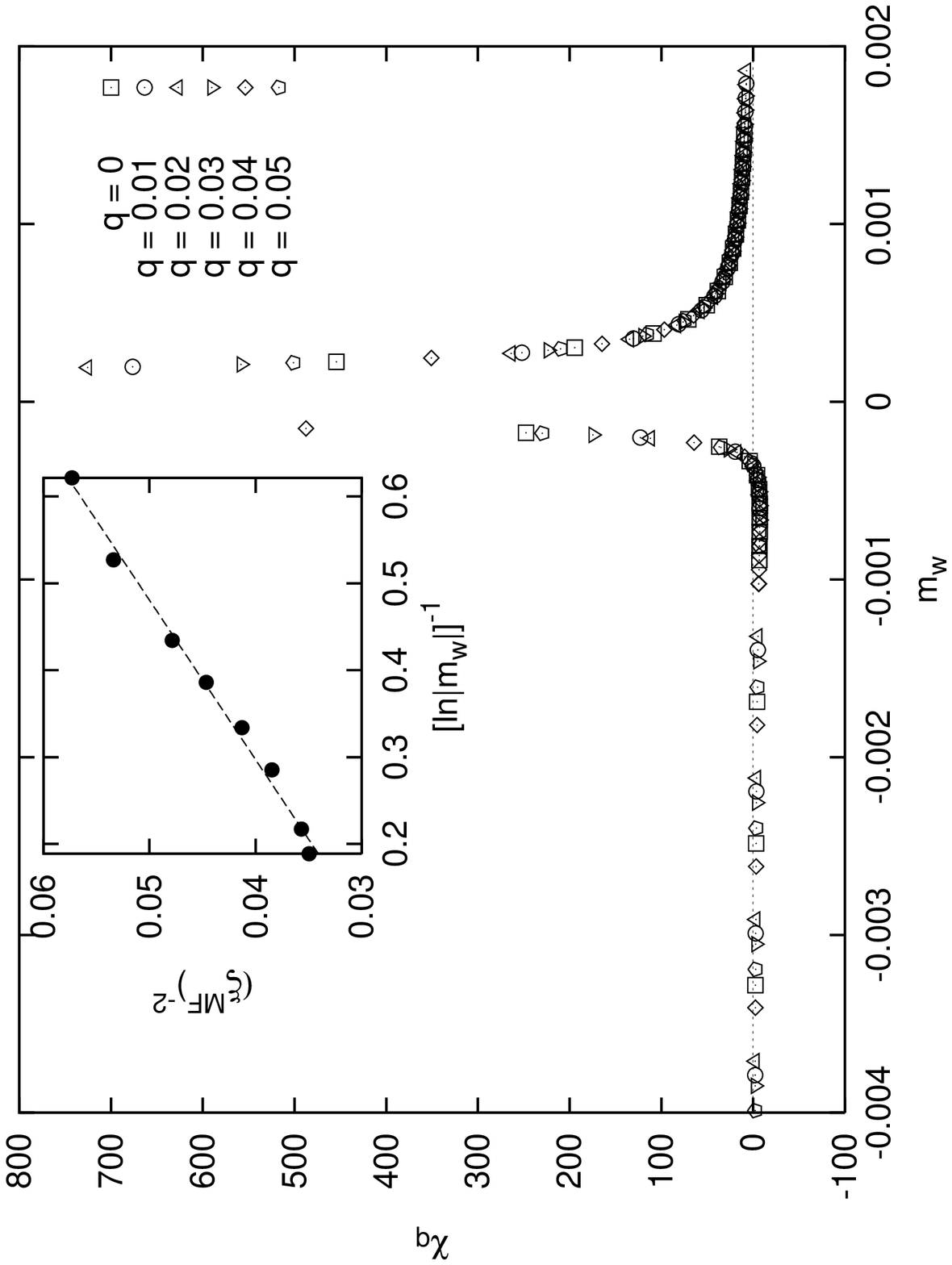}} \par}
\vspace{0.5001cm}

{\par\noindent \centering \textbf{Figure 5}.\par}
\vspace{0.5001cm}\newpage

\vspace{0.5001cm}
{\par\centering \resizebox*{0.7\textwidth}{0.7\textheight}{\includegraphics{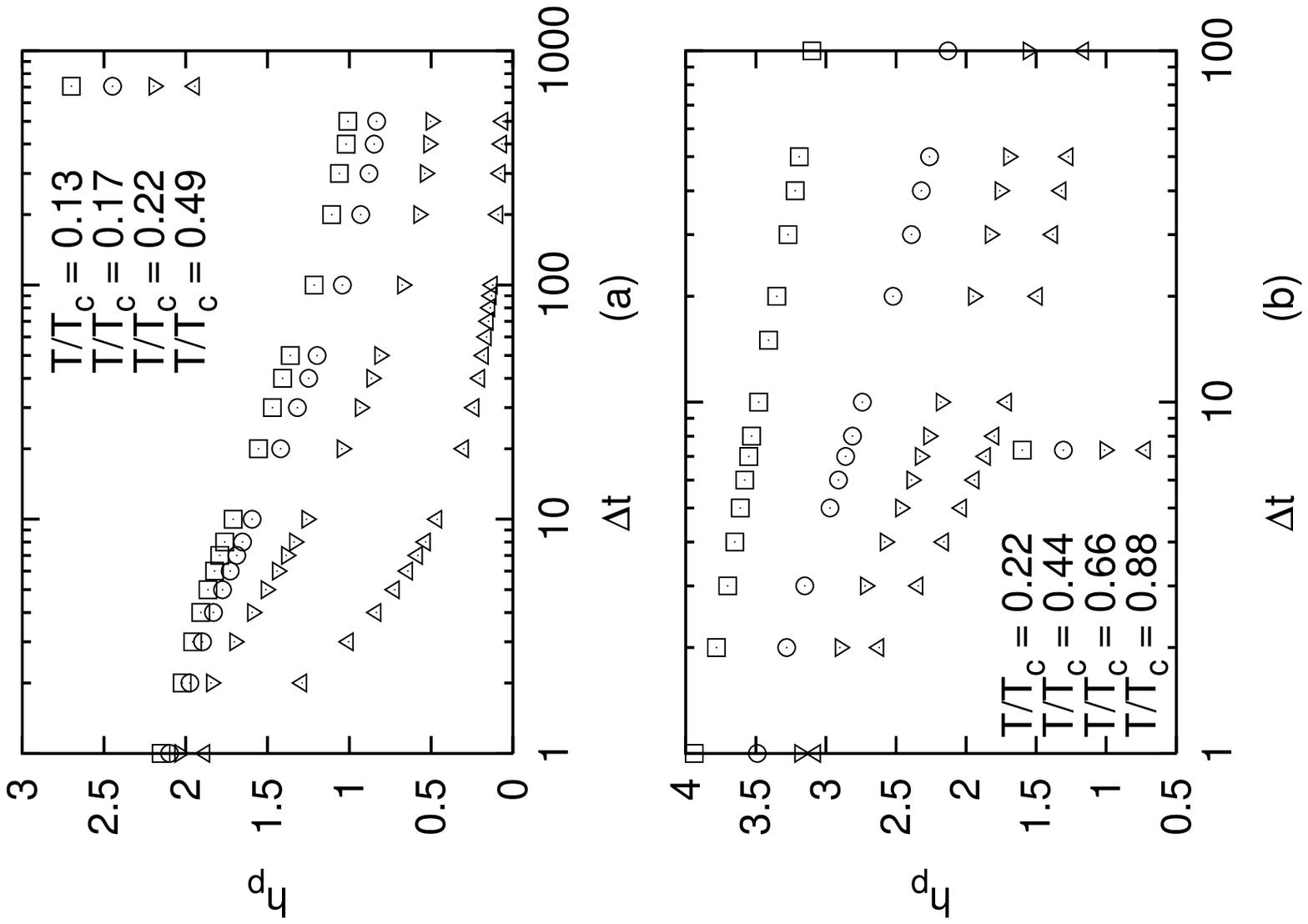}} \par}
\vspace{0.5001cm}

{\par\noindent \centering \textbf{Figure 6.}\par}
\vspace{0.5001cm}\newpage

\vspace{0.5001cm}
{\par\centering \resizebox*{0.7\textwidth}{0.7\textheight}{\includegraphics{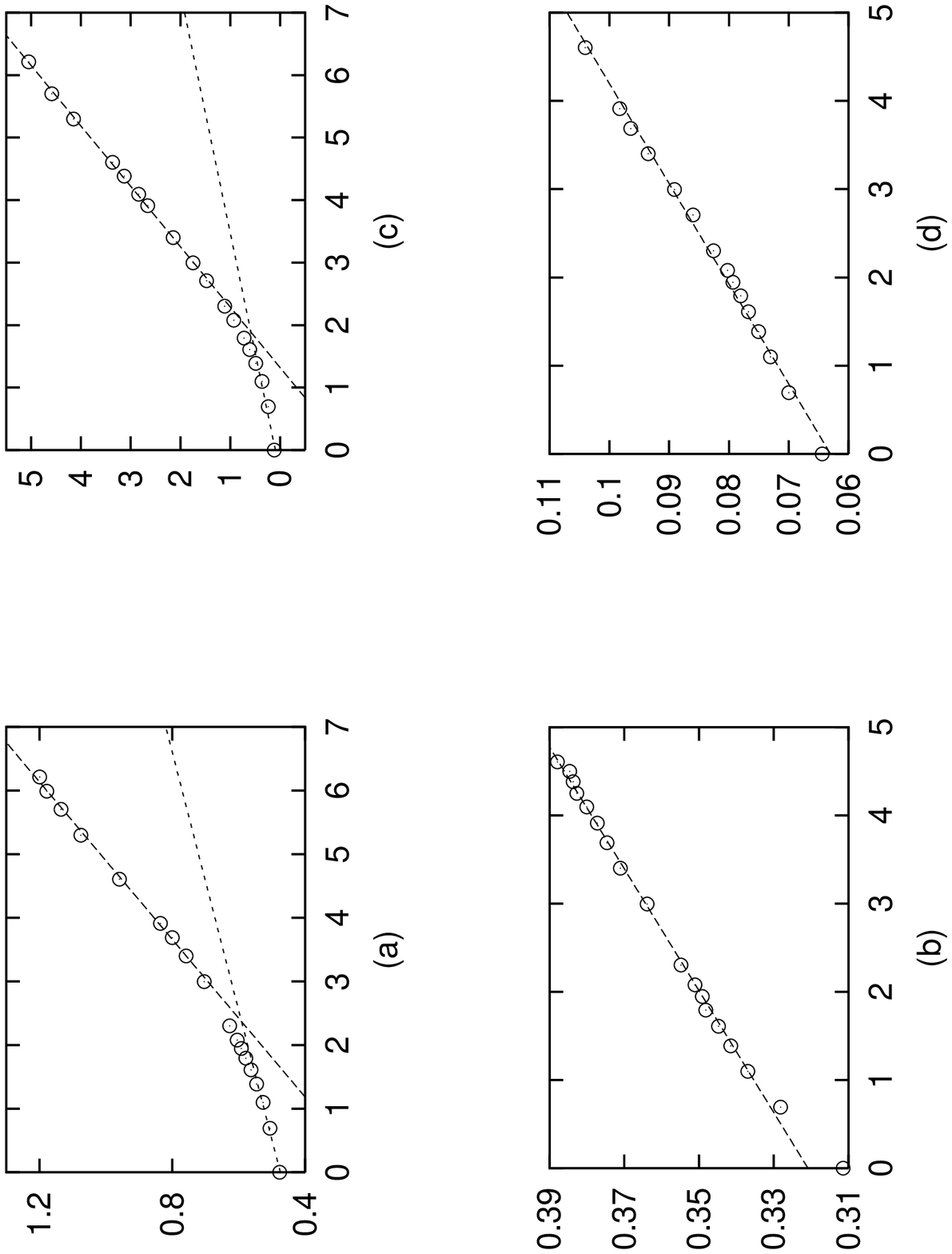}} \par}
\vspace{0.5001cm}

{\par\noindent \centering \textbf{Figure 7.}\par}
\vspace{0.5001cm}\newpage

\vspace{0.5001cm}
{\par\centering \resizebox*{0.7\textwidth}{0.7\textheight}{\includegraphics{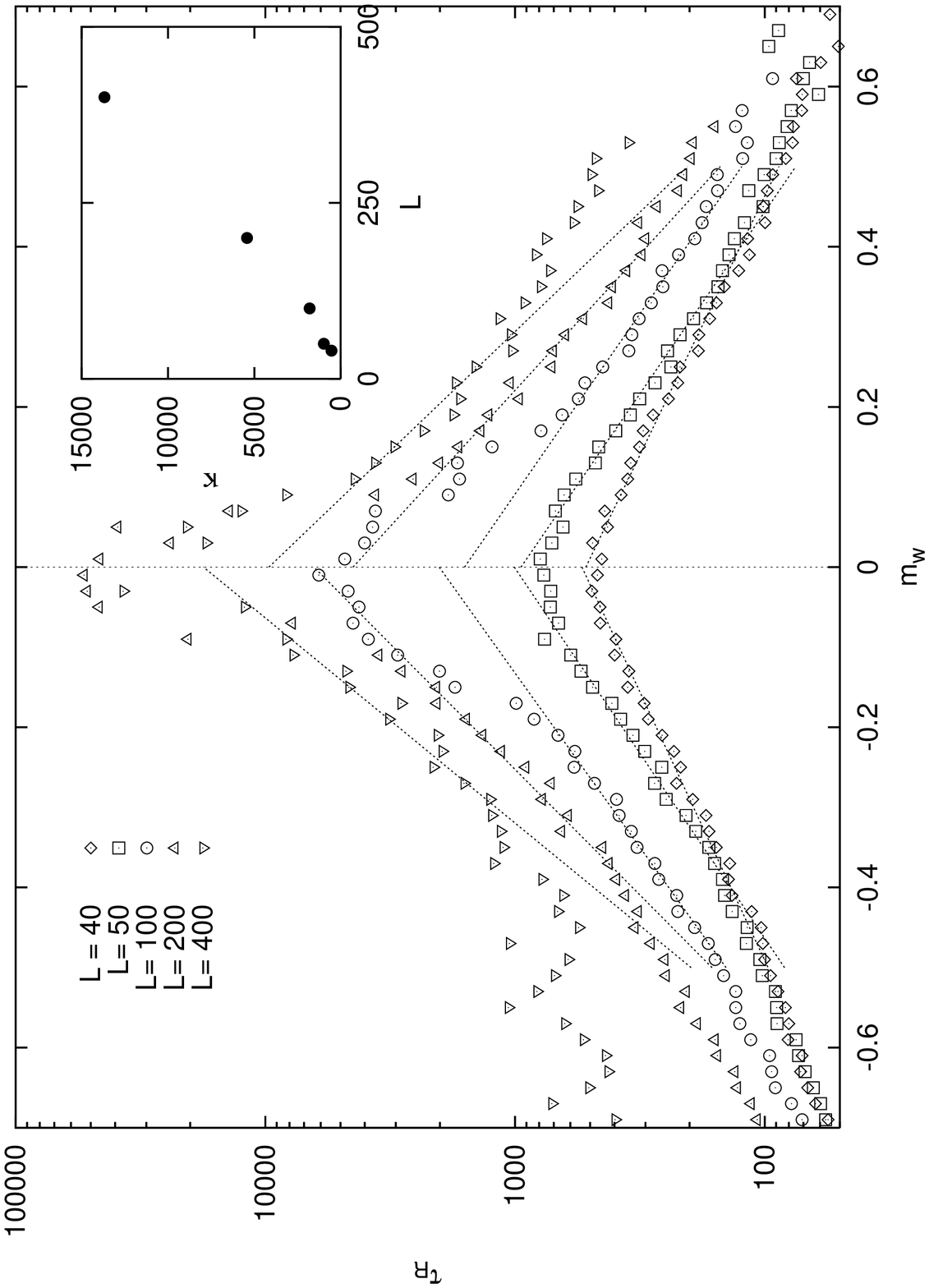}} \par}
\vspace{0.5001cm}

{\par\noindent \centering \textbf{Figure 8.}\par}
\vspace{0.5001cm}\newpage

\vspace{0.5001cm}
{\par\centering \resizebox*{0.7\textwidth}{0.7\textheight}{\includegraphics{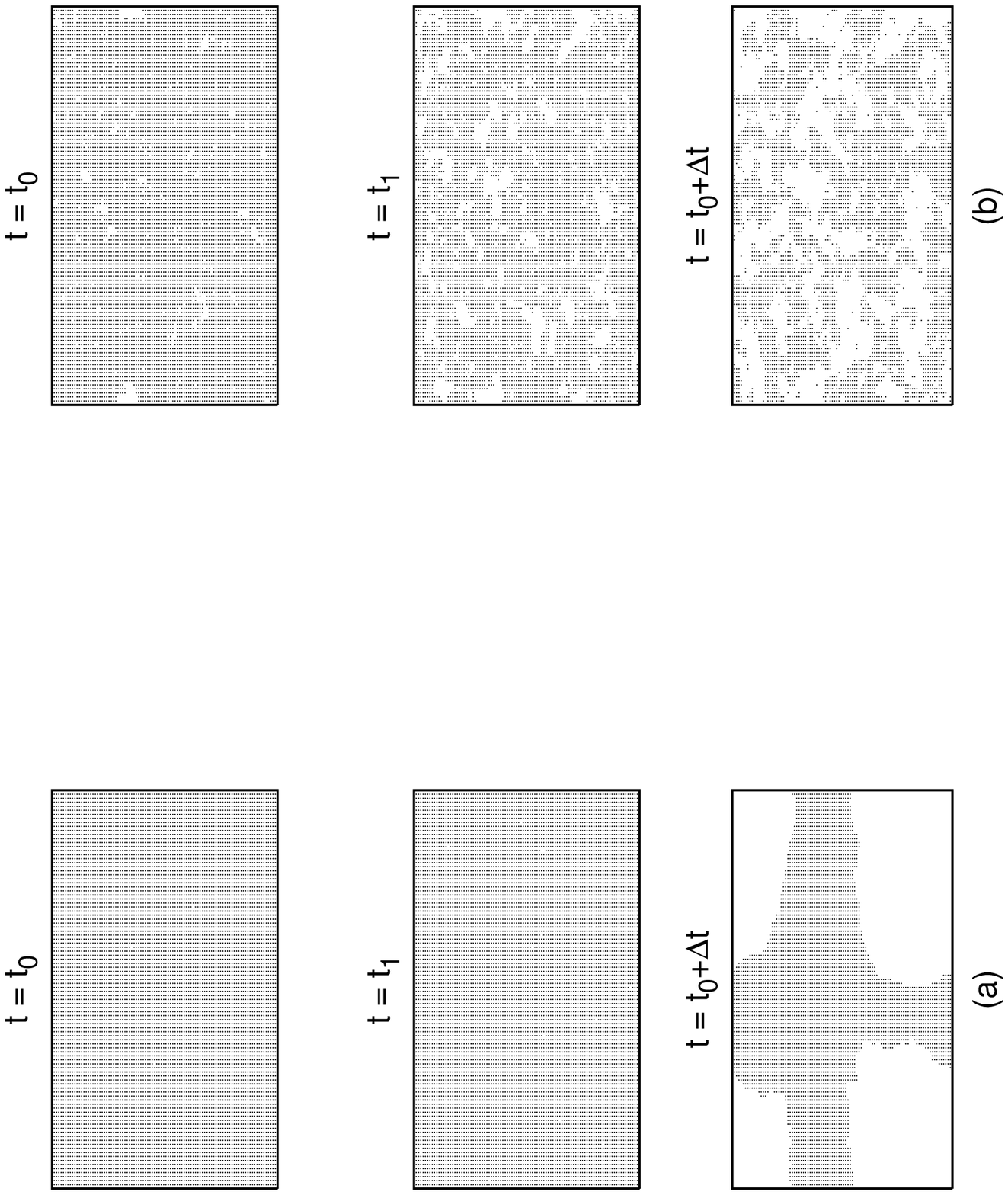}} \par}
\vspace{0.5001cm}

{\par\noindent \centering \textbf{Figure 9.}\par}
\vspace{0.5001cm}\newpage

\vspace{0.5001cm}
{\par\centering \resizebox*{0.7\textwidth}{0.7\textheight}{\includegraphics{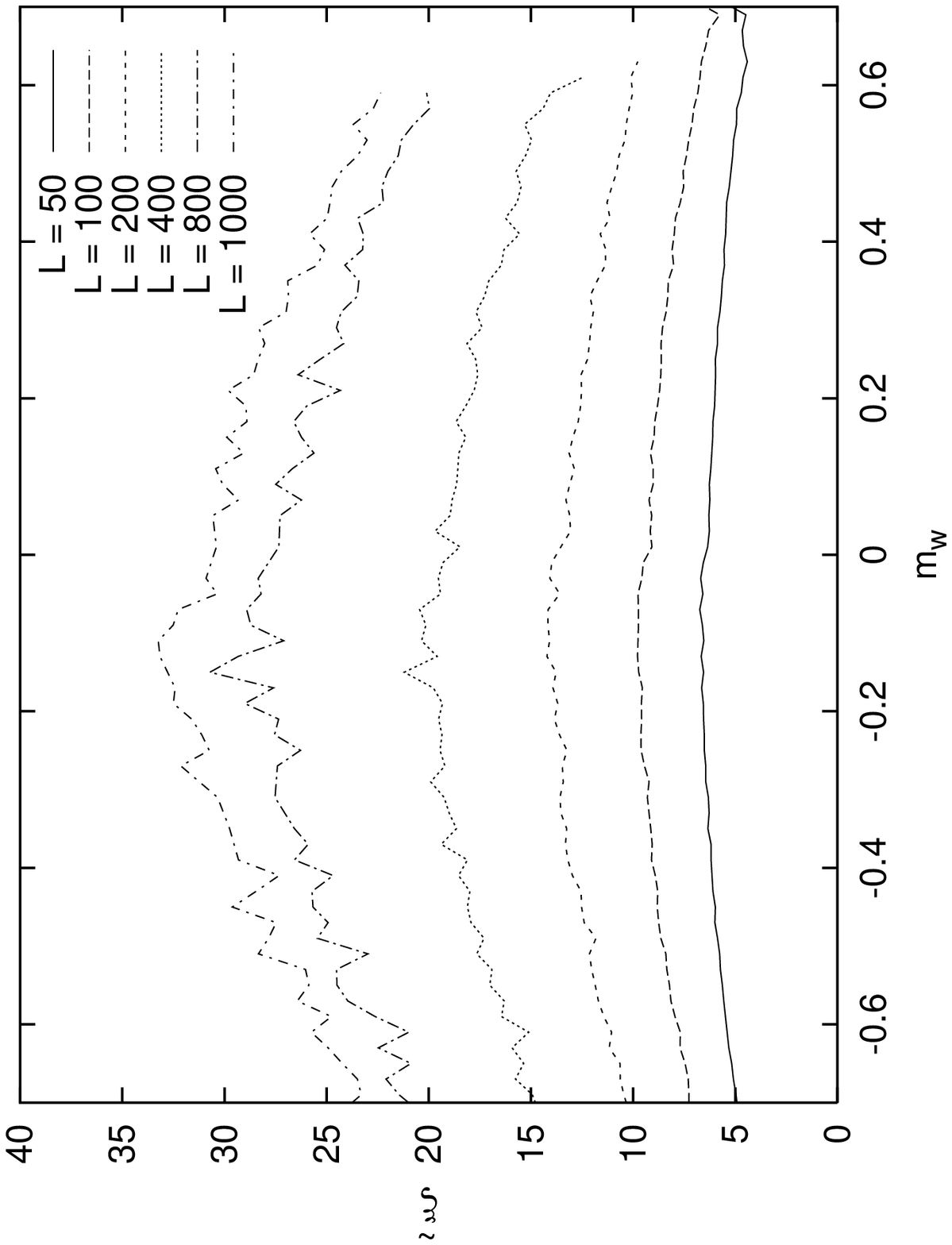}} \par}
\vspace{0.5001cm}

{\par\noindent \centering \textbf{Figure 10.}\par}
\vspace{0.5001cm}\newpage

\vspace{0.5001cm}
{\par\centering \resizebox*{0.9\textwidth}{0.5\textheight}{\includegraphics{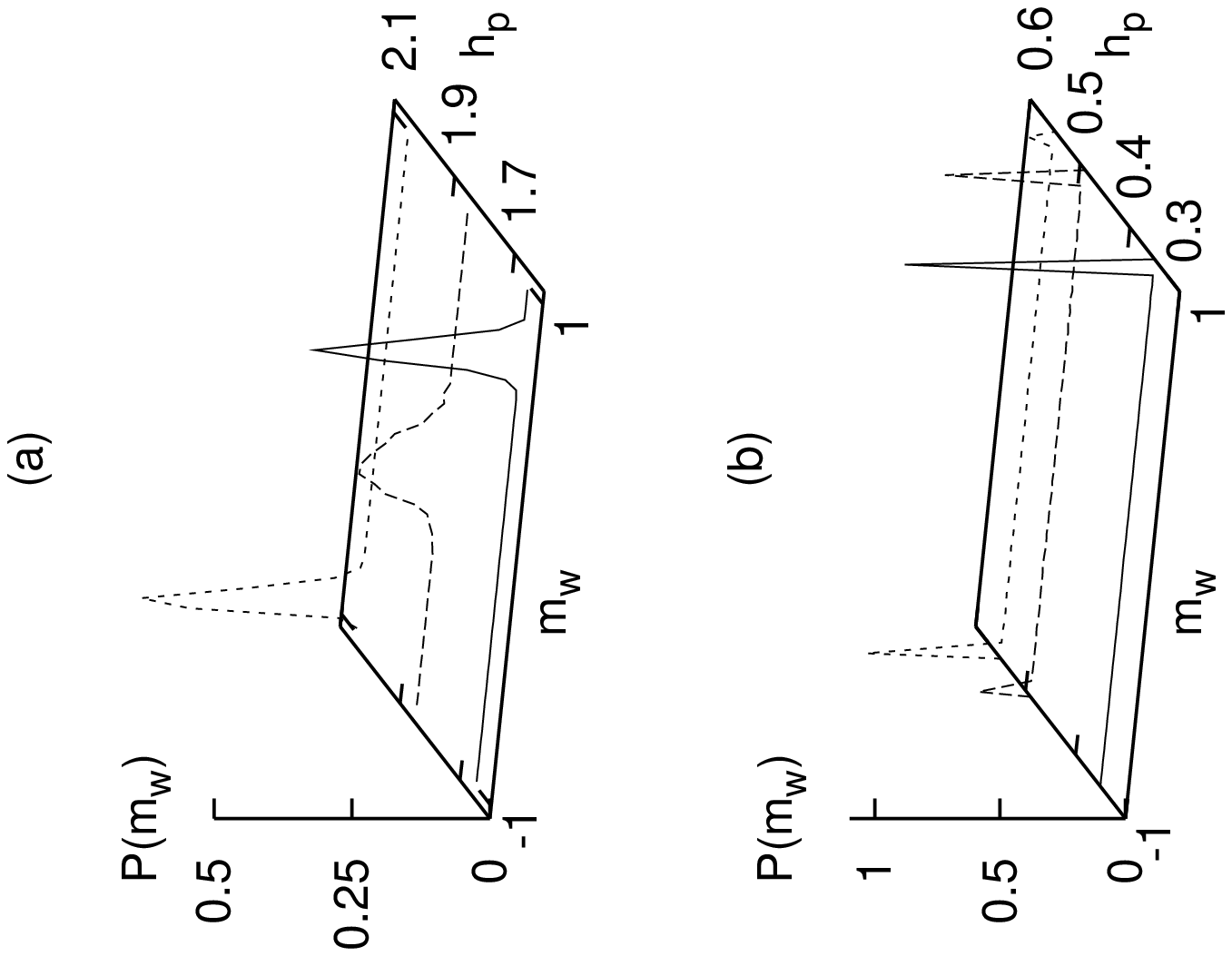}} \par}
\vspace{0.5001cm}

{\par\noindent \centering \textbf{Figure 11.}\par}
\vspace{0.5001cm}\newpage

\vspace{0.5001cm}
{\par\centering \resizebox*{0.75\textwidth}{0.4\textheight}{\includegraphics{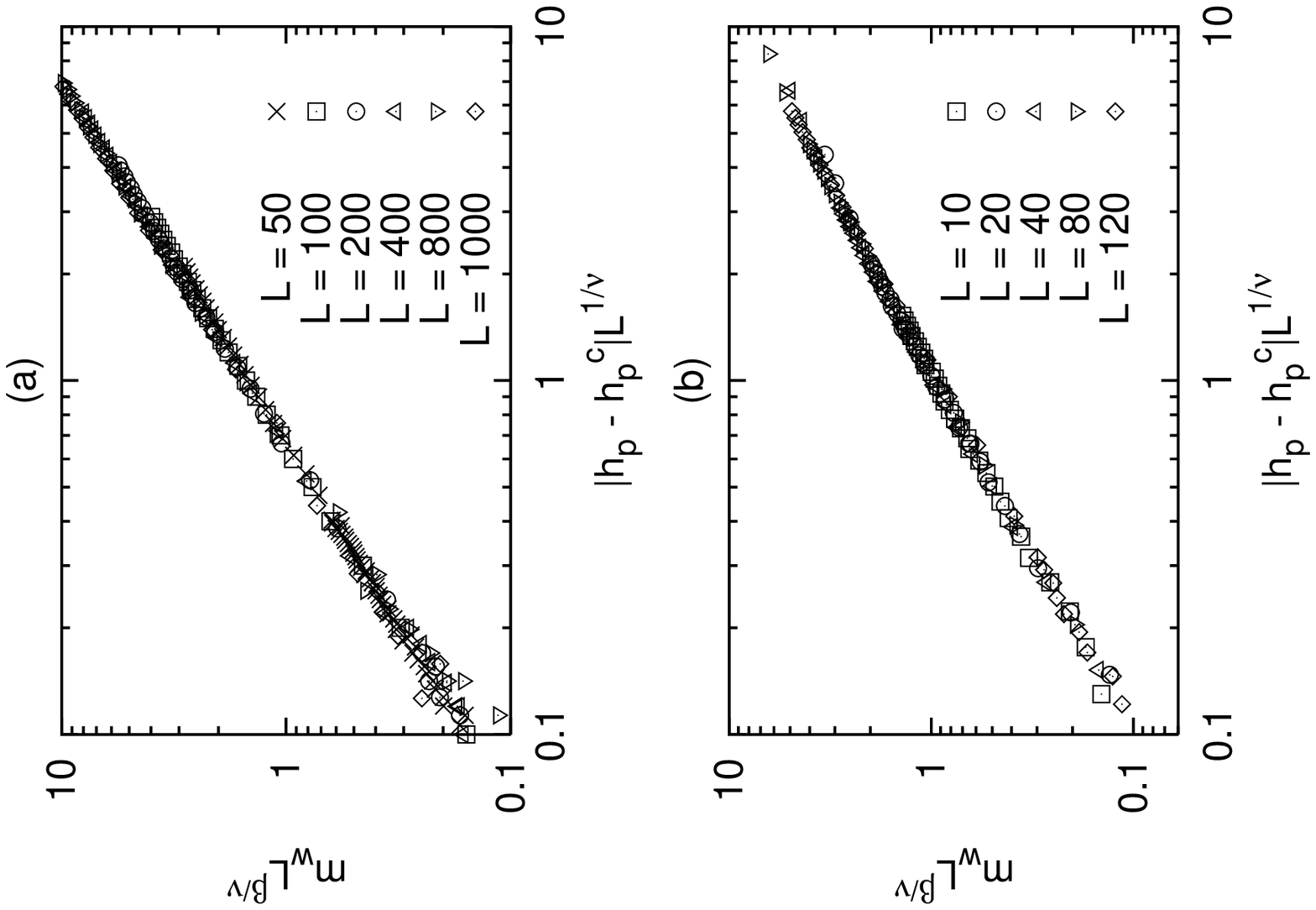}} \par}
\vspace{0.5001cm}

{\par\noindent \centering \textbf{Figure 12.}\par}
\vspace{0.5001cm}\newpage

\vspace{0.5001cm}
{\par\centering \resizebox*{0.7\textwidth}{0.7\textheight}{\includegraphics{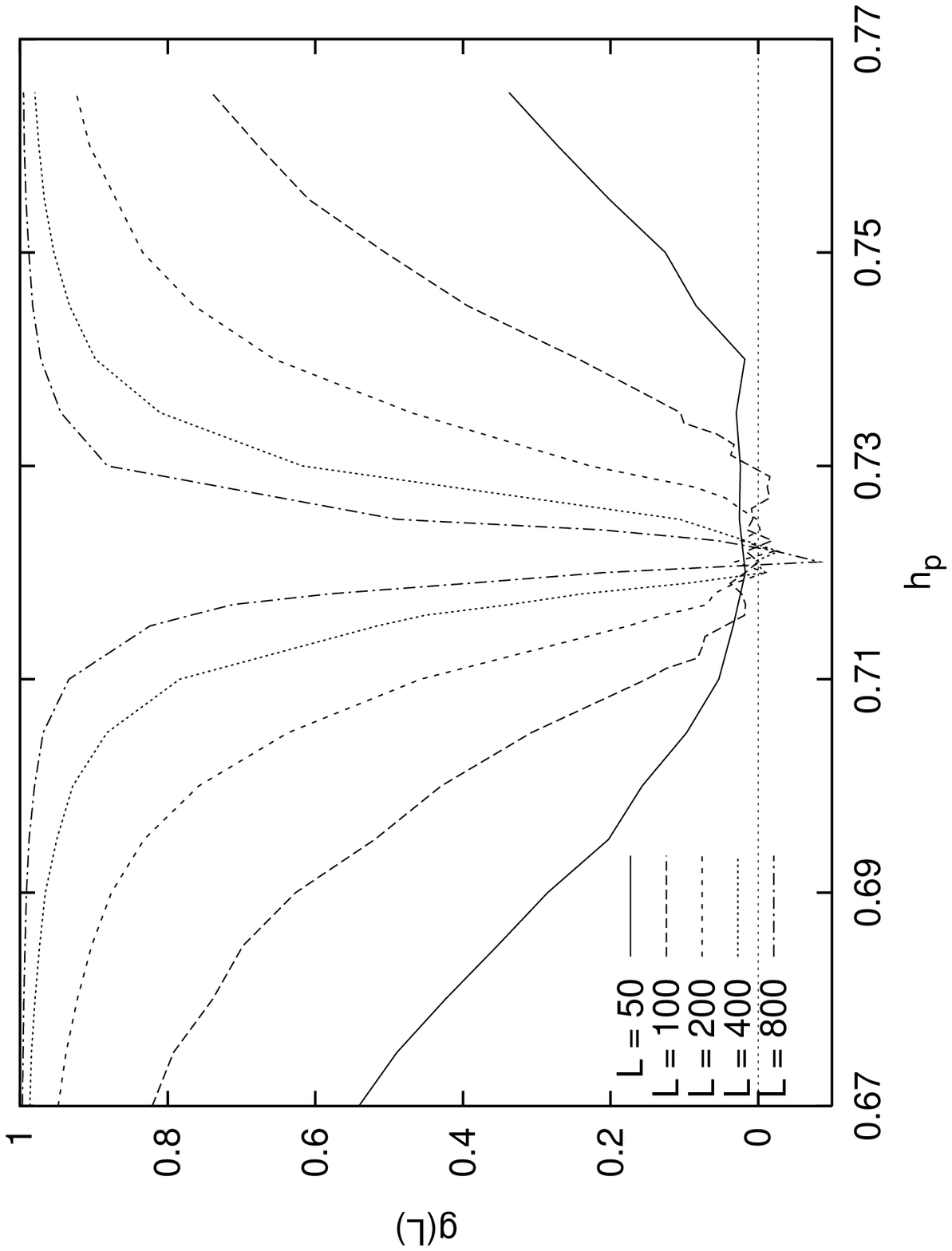}} \par}
\vspace{0.5001cm}

{\par\centering \textbf{Figure 13.}\par}


\begin{thebibliography}{}
\bibitem{ca99}Chakrabarti B K and Acharyya M 1999 \emph{Rev. Mod. Phys}. \textbf{71} 847
\bibitem{srn98.a}Sides S W, Rikvold P A and Novotny M A 1998 \emph{Phys. Rev}. E \textbf{57}
6512
\bibitem{mc98}Misra A and Chakrabarti B K 1998 \emph{Phys. Rev}. E \textbf{58} 4277
\bibitem{to90}Tome T and de Oliveira M J 1990 \emph{Phys. Rev}. A \textbf{41} 4251
\bibitem{ac95}Acharyya M and Chakrabarti B K 1995 \emph{Phys. Rev}. B \textbf{52} 6550
\bibitem{srn98.b}Sides S W, Rikvold P A and Novotny M A 1998 \emph{Phys. Rev}. \emph{Lett.} \textbf{81}
834
\bibitem{srn99}Sides S W, Rikvold P A and Novotny M A 1999 \emph{Phys. Rev}. E \textbf{59}
2710
\bibitem{abc97}Acharyya M, Bhattacharjee J K and Chakrabarti B K 1997 \emph{Phys. Rev}. E \textbf{55}
2392
\bibitem{mc97}Misra A and Chakrabarti B K 1997 \emph{Physica} A \textbf{247} 510
\bibitem{smc99}Stinchcombe R B, Misra A and Chakrabarti B K 1999 \emph{Phys. Rev}. E \textbf{59}
R4717
\bibitem{gd83}See e.g., Gunton J D and Droz M 1983 \emph{Introduction to the Theory of Metastable
and Unstable States} (Heidelberg : Springer)
\bibitem{sk68}Suzuki M and Kubo R 1968 \emph{J. of Phys. Soc. Japan} \textbf{24} 51
\bibitem{binder84}See e.g., Binder K 1984 \emph{Application to the Monte-Carlo Method in Statistical
Physics} (Heidelberg : Springer)
\bibitem{blh95}Bl\( \ddot{\textrm{o}} \)te H W J, Luijten E and Heringa J R 1995 \emph{J.
Phys.} A \textbf{28} 6289
\bibitem{rtms94}Rikvold P A, Tomita H, Miyashita S and Sides S W 1994 \emph{Phys. Rev}. E \textbf{49}
5080
\bibitem{as98}Acharyya M and Stauffer D 1998 \emph{Euro. Phys. J}. B \textbf{5} 571 
\bibitem{mc00}Misra A and Chakrabarti B K 2000 arXiv:cond-mat/0002085
\bibitem{robin85}Stinchcombe R B 1985 \emph{Scaling Phenomena in Disordered Systems} ed. Pynn
R and Skeljorp A (New York : Plenum)
\bibitem{barber83}See e.g., Barber M N 1983 \emph{Phase Transition and Critical Phenomena} eds.
Domb C and Lebowitz J L (New York : Academic)
\bibitem{binder88}See e.g., Binder K and Heermann D 1988 \emph{Monte Carlo Simulation in Statistical
Physics} (Berlin : Springer-Verlag)
\bibitem{rkwns00}Rikvold P A, Korniss G, White C J, Novotny M A and Sides S W 2000 \emph{Computer
Simulation Studies in Condensed Matter Physics} XII eds. Landau D P et. al.
(Berlin : Springer)
\end{thebibliography}
\end{document}